\documentclass[12pt]{article}
\usepackage{graphicx}
\usepackage{wrapfig}
\usepackage{epsfig}
\usepackage{epsfig}
\usepackage{graphics}
\usepackage{epic}

\newcommand\pubnumber{XXX}
\newcommand\pubdate{\today}

\def\isu{Department of Physics and Astronomy\\
Iowa State University, Ames, Iowa, USA}

\def\Title#1{\begin{center} {\Large #1 } \end{center}}
\def\Author#1{\begin{center}{ \sc #1} \end{center}}
\def\Address#1{\begin{center}{ \it #1} \end{center}}

\newcommand\pubblock{\rightline{\begin{tabular}{l} \pubnumber\\
         \pubdate  \end{tabular}}}
\newenvironment{Abstract}{\begin{quotation}  }{\end{quotation}}
\newenvironment{Presented}{\begin{quotation} \begin{center} 
             PRESENTED AT\end{center}\bigskip 
      \begin{center}\begin{large}}{\end{large}\end{center} \end{quotation}}

\input babarsym

\begin{document}
\begin{titlepage}
\pubblock

\vfill
\Title{Time-dependent \g/$\phi_3$ measurements by \babar}
\vfill
\Author{Ada E. Rubin\\
on behalf of the \babar\ Collaboration}
\Address{\isu}
\vfill
\begin{Abstract}
Compilation and summary of 
time-dependent measurements
of the CKM angle $\gamma/\phi_3$
with events collected at the 
$\babar$ detector at the SLAC PEP-II asymmetric $B$ factory.
\end{Abstract}
\vfill
\begin{Presented}
CKM 2010, the 6th International Workshop on CKM\\
Unversity of Warwick, UK\\
August 6--10, 2010
\end{Presented}
\vfill
\end{titlepage}

\section*{Introduction}

An important goal of flavor physics 
is to overconstrain the CKM elements. 
The CKM element $\gamma/\phi_3$ is the least precisely
measured of the Unitarity Triangle angles.
Decays of $B_d$ mesons that allow one to constrain 
the CKM angle $\rm{sin}(2\beta+\gamma)$
have either small $CP$ asymmetry ($B\to D^{(*)}\pi/\rho$ and $\Bz\to D^{\mp} \KS \pi^{\pm}$) 
or small branching fractions ($B \to D^{(*)} K^{(*)}$). 
The $CP$ violating effects in these modes, therefore, are difficult to measure.

The quantity $\rm{sin}(2\beta+\gamma)$ can be obtained 
from the 
study of the time evolution of 
$\Bz/\Bzb \to D^{(*)} X_{u,d,s}$ decays 
where $X_{u,d,s}$ refers to light and/or strange mesons.
In the Standard Model, these decays
proceed via Cabibbo suppressed $\b \to u$ and
favored $\b\to c$ transitions  
described by the amplitudes $A_u$ and $A_c$, respectively.
The magnitude of the ratio between the amplitudes $A_u$ and 
$A_c$ is $r$.
%
The relative weak phase between these two amplitudes
is $\gamma$; it is  $2\beta+\gamma$ with $\Bz \Bzb$ mixing.
Also, there exists the strong phase difference
between these two amplitudes, $\delta$.
These hadronic parameters 
in the observables,  $r$ and $\delta$, 
make extraction of the weak phase information difficult.



The time dependent (TD) distribution for $\Bz$ decays to a final state 
can be written as
\begin{eqnarray}
f^{\pm}&=&
 \frac{e^{-\left|\Delta t\right|/\tau}}{4\tau} \times 
[ 1 \mp 
 S^{\pm}_\eta \sin(\Delta m_d \Delta t) \mp
  \eta C \cos(\Delta m_d \Delta t)]
\label{td-eq}
\end{eqnarray}
where $\tau$ is the $\Bz$ lifetime, 
$\Delta m_d$ is the $\BzBzb$ mixing frequency and 
$\Delta t = t_{\rm rec} - t_{\rm tag}$ 
is the time of the reconstructed $B$  ($B_{\rm rec}$)  decay 
relative to the decay of the other $B$ 
($B_{\rm tag}$) from the $\FourS \rightarrow \BB$ decay. 
$\Delta t$ is calculated from the measured  separation along the
beam collision axis ($z$)
between the $B_{\rm rec}$ and $B_{\rm tag}$ decay vertices: 
$\Delta z$=$\beta\gamma c \Delta t$ where $\beta\gamma$=0.56
is the  Lorentz boost of $\BB$ pairs along the direction of the high-energy beam.
In equation~\ref{td-eq} the upper (lower) sign refers to the flavor
 of $B_{\rm tag}$ as $\Bz$ $(\Bzb)$,
while $\eta=+1$ ($-1$) denotes the final state
$D^{(*)}$ ($\bar{D^{(*)}}$).
The specifics of the
$CP$ parameters, $S^{\pm}_\eta$ and $C$, depend on the physics of the
reconstructed \Bz decay mode.


\section*{$CP$ asymmetry in $\Bz \rightarrow D^{(*)\mp} \pi^{\pm}/\rho^\pm$ decays}

The decay modes $\Bz \rightarrow D^{(*)\mp} \pi^{\pm}$ have been
proposed  to measure
$\sin(2\beta+\gamma)$~\cite{GammaFromDpi}.
The decay rate distribution for 
$B \to D^{(*)\mp}\pi^\pm$ is given by equation~\ref{td-eq}
which is parametrized to account for tag-side interference~\cite{DpiParam}.
The $CP$ parameter $C$ is unity  and $S^\pm$ for each tagging category is given by 
$S^{\pm}_{\eta} = (a -  \eta c)$ with
  $a = 2 r \sin(2\beta+\gamma) \cos \delta$, 
  $c = 2 \cos(2\beta+\gamma) (r\sin \delta)$.
Since $A_u$ is doubly CKM-suppressed with respect
to $A_c$, one expects the ratio to be of order 2\%.
Due to the small value of $r$, large data samples
are required for a statistically significant measurement of $S^{\pm}_{\eta}$. 


Fully reconstructed 
$\Bz \rightarrow D^{(*)\mp} \pi^{\pm}$ and  
$\Bz \rightarrow D^{\mp} \rho^{\pm}$  decays~\cite{BtoDpiFull} using 232 million \BB pairs
are used to measure the parameters $a$ and $c$.
Results of this analysis from the TD maximum likelihood fit 
are 
\begin{eqnarray} 
  a^{D\pi}              =& -0.010\pm 0.023  \,\pm 0.007 \, ,~c_{\rm lep}^{D\pi}    &= -0.033\pm 0.042  \,\pm 0.012 \,\,  \nonumber \\
  a^{D^*\pi}           =& -0.040\pm 0.023  \,\pm 0.010 \, ,~c_{\rm lep}^{D^*\pi}  &=  \phantom{-}0.049\pm0.042  \,\pm 0.015 \,\, \nonumber \\
  a^{D\rho}             =& -0.024\pm 0.031  \,\pm 0.009 \, ,~c_{\rm lep}^{D\rho}   &= -0.098\pm 0.055  \,\pm 0.018 \,\, \nonumber 
\end{eqnarray}
where the first error is statistical and the second is systematic. 

\begin{wrapfigure}{r}{0.48\textwidth}
\vspace{-20pt}
  \begin{center}
    \includegraphics[width=0.48\textwidth]{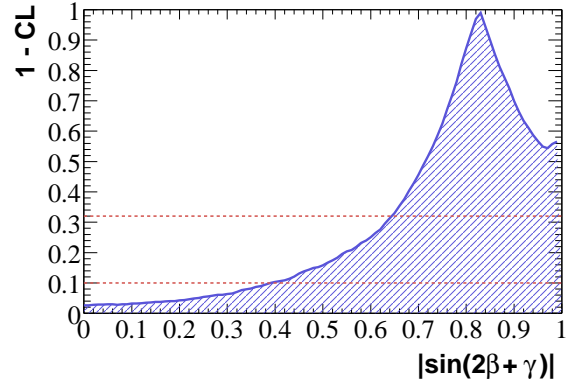}
  \end{center}
  \vspace{-20pt}
  \caption{\label{cl-rs-sin2betapg}{\it{The shaded region denotes the allowed range of $|\sin(2\beta+\gamma)|$
for each confidence level. The horizontal lines show, from top
to bottom, the 68\% and 90\% CL.}}}
\vspace{-10pt}
\end{wrapfigure}

In partially reconstructing   $\Bz\to D^{*\mp}\pi^{\pm}$ candidates,
only the hard (high-momentum) pions $\pi_h$ from $B$ decay and 
soft (low-momentum) pions $\pi_s$ from 
$D^{*-}\rightarrow \Dzb \pi_s^-$ decays are employed.
The  ``missing mass``  of  the non-reconstructed 
$D$ is the kinematic variable used to extract 
signal events; it peaks 
at the nominal $\Dz$ mass.
This method eliminates the efficiency loss associated with 
\Dz meson reconstruction. The $CP$ asymmetry measured with this technique~\cite{BtoDpiPartial}
using 232 million \BB pairs is 
\begin{eqnarray*}
a^{D^*\pi}  =& -0.034\pm0.014\pm 0.009 \,  ,\\
~c_{\rm lep}^{D^*\pi}  =& -0.019\pm0.022\pm 0.013 \,
\label{math:dstpi_partial}
\end{eqnarray*}


To interpret these results in terms of constraints on $|\sin(2\beta+\gamma)|$ ,
findings from the  fully reconstructed  $\Bz \rightarrow D^{(*)\mp} \pi^{\pm}$ ,
$\Bz \rightarrow D^{\mp} \rho^{\pm}$ analysis 
are combined with those of the
partially reconstructed   $\Bz\to D^{*\mp}\pi^{\pm}$ study
using a frequentist method described in Ref.~\cite{BtoDpiPartial}.
This method sets the lower limits 
$|\sin(2\beta+\gamma)|\!>\!0.64\ (0.40)$ at $68\%$ $(90\%)$ C.L.  as seen in Figure~\ref{cl-rs-sin2betapg}.

\section*{Dalitz plot analysis of $\Bz\to D^{\mp} \Kz \pi^{\pm}$}

Measurement of $\sin(2\beta+\g)$ from
three body $B$ decays, such as $\Bz\to D^{\mp} \Kz \pi^{\pm}$ have been suggested as a way to avoid the limitation 
of small $r$, since $r$ in these decays could be as large as 0.4 in some regions of the Dalitz
plane~\cite{theo-Dalitz}. The final state, $D^{\mp} \Kz \pi^{\pm},~\Dp \to \Km\pip\pim$, is reached 
via the following intermediate states: 
$\Bz\to D^{**0} \KS$ with $D^{**0}=\{D_0^{**}(2400),~D_2^{**}(2460)\}$ , 
$\Bz \to \Dm \Kstarp$ with  $K^* = \{K^*(892),~K^*_0(1430),~K^*_2(1430),~K^*(1680)\}$,
and a small expected contribution from $\Bz \to D_s^{*+}(2573)\pi^-$.
 The TD Dalitz plot PDF is of the same form as equation~\ref{td-eq}, but
multiplied by the factor $(A_c^2+A_u^2)/2$ and with the coefficient of the $\sin$ term being 
\[ S_\eta = \frac{2 \mathrm{Im} (A_{c}A_{u} e^{i(2\beta+\g)+\eta i(\phi_{c}-\phi_{u})})}{A_{c}^2+A_{u}^2}. \]
The amplitudes ($A_{c},~A_{u}$) and strong phases ($\phi_{c}, \phi_{u} $)
are functions of their positions in the Dalitz plot.
The coefficient of the $\sin$ term is $C = (A_{c}^2-A_{u}^2)/ (A_{c}^2+A_{u}^2)$.

 \begin{wrapfigure}{r}{0.5\textwidth}
 \vspace{-45pt}
 \begin{center}
    \includegraphics*[width=0.48\textwidth]{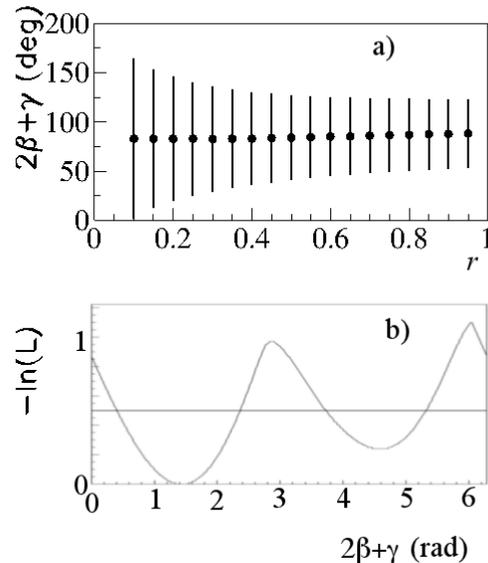}
  \end{center}
  \vspace{-20pt}
\caption{ \it a): distribution of the values of $2\beta+\g$ fitted on
data for different
hypotheses on the $r$ value. b): variation of the logarithm of the
likelihood with $2\beta+\g$.} 
\label{fig-BtoDKSpi}
\vspace{-40pt}
\end{wrapfigure}

With the ratio of the amplitudes $r$ set to  0.3 for each resonance in the PDF,
consistent with the limit $r<0.4$ (90\%  CL) reported in Ref.\cite{BtoDKSBabar},
the weak phase  is found to be 
$2\beta+\g = (83\pm53\pm20)^\circ$ and $(263\pm53\pm20)^\circ$~\cite{BtoDKpiDalitzPlot}, shown
in Fig.~\ref{fig-BtoDKSpi}b,
in a sample of  347 million \BB pairs.
The central value $2\beta+\g$ is stable with respect 
to the value of $r$ (Fig.~\ref{fig-BtoDKSpi}a).


\section*{$\Bzb\to D^{(*)0}\bar{K}^{0}$ decays}

The decay modes $\Bzb\to D^{(*)0}\bar{K}^0$ have been proposed for determination 
of $\sin(2\beta+\gamma)$ from measurement of TD \CP asymmetries~\cite{ref:th-dst0k0}. 
Due to relatively large \CP asymmetry 
($r_B\equiv|A(\Bzb\to \bar{D}^{(*)0}\bar{K}^0)|/|\Bzb\to D^{(*)0}\bar{K}^0)|\simeq 0.4$) 
these decays appear ideal for such a measurement. 
The TD decay rate in this case can be parameterized such that
$C=(1-r_B^2)/(1+r_B^2)$  and $S=r_B \sin(2\beta+\g+\delta) / (1+r_B^2)$.
Since $r_B$ 
can simply be measured by fitting the $C$ coefficient  in the decay distributions, 
the measured asymmetry can be interpreted in terms of $\sin(2\beta+\gamma)$ without additional assumptions. 
However, the branching fractions of such decays are relatively small, $\mathcal{O}(10^{-5})$. 
Therefore a large data sample is required. 

\begin{figure}[htb]
 \vspace{-10pt}
 \begin{center}
  \includegraphics[width=0.4\linewidth]{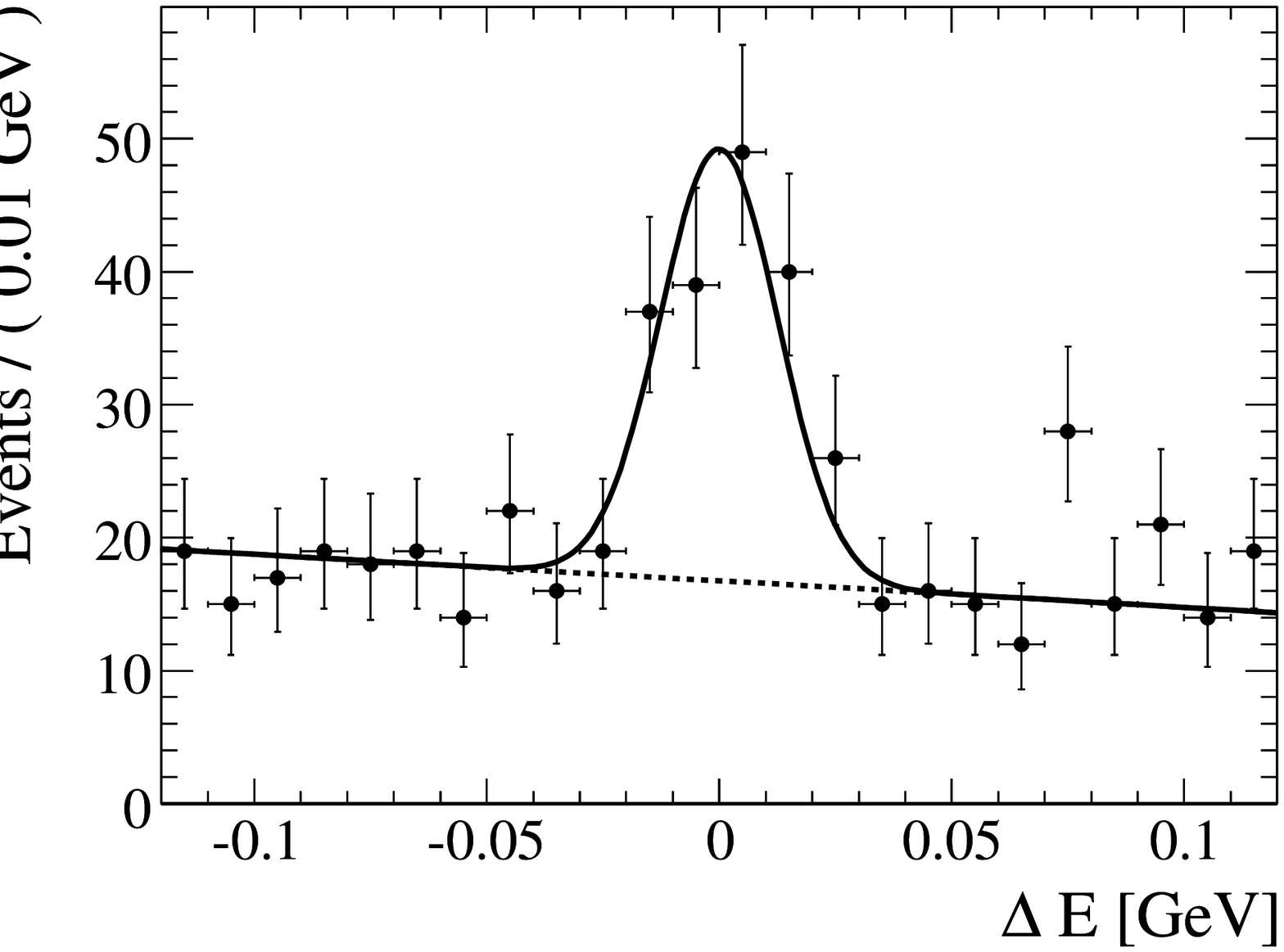}
  \put(-125,95){a)}%
  \includegraphics[width=0.4\linewidth]{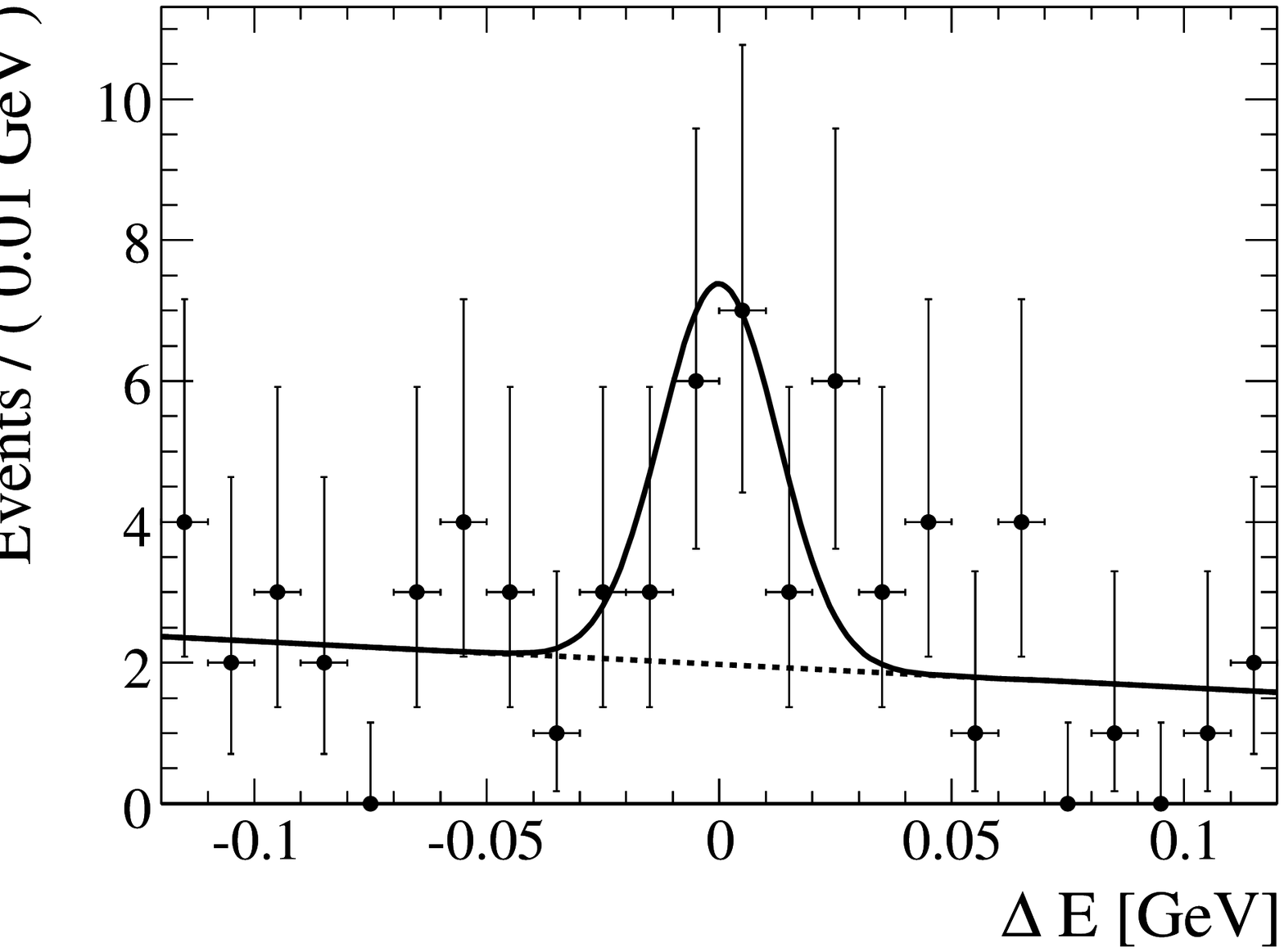}
  \put(-125,95){b)}\\
  %
   \vspace{-20pt}
\end{center}
  \caption{\label{fig:DK} \it
    Distribution of \DeltaE\ for  a) $\Bzb\to \Dz\Kzb$, b)
    $\Bzb\to \Dstarz\Kzb$, 
    The
    points are the data, the solid curve is the projection of the likelihood fit, and
    the dashed curve represents the background component.}
\end{figure}

The most recent measurement~\cite{BtoDKSBabar} of these decays  using a data sample of 226 million \BB pairs 
finds
\begin{eqnarray*}
  \BR(\Bzb\to \Dz\Kzb)     &=& (5.3\pm0.7\pm0.3)\times 10^{-5}\, \nonumber \\
  \BR(\Bzb\to \Dstarz\Kzb) &=& (3.6\pm1.2\pm0.3)\times 10^{-5}\,
\end{eqnarray*}
from signal yields to the maximum likelihood fits in Fig.~\ref{fig:DK}.
With just over 100 signal events, a TD decay rate analysis is not feasible. 

\section*{Conclusion}

Non-trivial, theoretically clean constraints on $2\beta+\gamma$ come from  
measurements of time-dependent \CP asymmetry in the $B$ decays.
Updated measurements to the full \babar\ dataset of 468 million \BB pairs
will only deepen our understanding of the CKM  mechanism.
We expect an improvement in the measurement of \g with
  $\B \to D^{(*)\mp} \pi^{\pm}/\rho^{\pm}$  since $r$ can be  more 
precisely estimated 
by using the isospin 
relation $r=\sqrt{\frac{\tau_\Bz}{\tau_\Bp} \, \frac{2\BR(\Bp \to D^{*+}\piz)}{\BR(\Bz \to D^{*-}\pi^+)}}<0.051$  (90\% C.L.) as suggested by Ref.~\cite{belle-thing}.
It is also possible that the full \babar\ data sample is just large enough to detect \CP asymmetry in the mode \Bzb\to \Dz\Kzb.

\end{document}